\def\ra{\rangle}
\def\la{\langle}
\def\be{\begin{equation}}
\def\ee{\end{equation}}
\def\ba{\begin{array}}
\def\ea{\end{array}}

\documentclass[aps,pre,preprint,showpacs,amsmath,amssymb,amsfonts,preprintnumbers]{revtex4}
\usepackage{epsfig,graphicx}

\begin{document}

\baselineskip=18pt \setcounter{page}{1} \centerline{\large\bf Bounds
for multipartite concurrence} \vspace{4ex}
\begin{center}
Ming Li$^{1}$, Shao-Ming Fei$^{2,3}$ and  Zhi-Xi Wang$^{2}$

\vspace{2ex}

\begin{minipage}{5in}

\small $~^{1}$ {\small College of Mathematics and Computational
Science, China University of Petroleum, 257061 Dongying, China}

\small $~^{2}$ {\small Department of Mathematics, Capital Normal
University, 100037 Beijing, China}

\small $~^{3}$ {\small Max-Planck-Institute for Mathematics in the
Sciences, 04103 Leipzig, Germany}

\end{minipage}
\end{center}

\begin{center}
\begin{minipage}{5in}
\vspace{1ex} \centerline{\large Abstract} \vspace{1ex} We study the
entanglement of a multipartite quantum state. An inequality between
the bipartite concurrence and the multipartite concurrence is
obtained. More effective lower and upper bounds of
the multipartite concurrence are obtained.
By using the lower bound, the entanglement of
more multipartite states are detected.

\smallskip
PACS numbers: 03.67.-a, 02.20.Hj, 03.65.-w\vfill
\smallskip
\end{minipage}\end{center}
\bigskip
As a potential resource for communication and
information processing,
quantum entanglement has rightly been the subject of much study in
recent years \cite{nielsen}. However the boundary between
the entangled states and the separable states, states that can be
prepared by means of local operations and classical communications
\cite{werner}, is still not well characterized. Entanglement
detection turns out to be a rather tantalizing problem. A more
general question is to calculate the well defined quantitative
measures of quantum entanglement such as entanglement of formation
(EOF) \cite{eof} and concurrence \cite{con,multicon}. A series of excellent
results have been obtained recently.

There have been some (necessary) criteria for separability, the Bell
inequalities \cite{bell64}, PPT (positive partial transposition)
\cite{peres} (which is also sufficient for the cases $2\times2$ and
$2\times3$ bipartite systems \cite{2231}), realignment
\cite{ChenQIC03,ru02,chenPLA02} and generalized realignment
\cite{chenkai}, as well as some necessary and sufficient operational
criteria for low rank density matrices\cite{hlvc00,afg01,feipla02}.
Further more, separability criteria based on local uncertainty
relation \cite{hofmann,0604050,117903,0611282} and the correlation
matrix \cite{julio,hassan} of the Bloch representation for a quantum
state have been derived, which are strictly stronger than or
independent of the PPT and realignment criteria. The calculation of
entanglement of formation or concurrence is complicated except for
$2\times 2$ systems \cite{wootters} or for states with special forms
\cite{conspacial}. For general quantum states with higher dimensions
or multipartite case, it seems to be a very difficult problem to
obtain analytical formulas. However, one can try to find the lower and
the upper bounds to estimate the exact values of the concurrence
\cite{chenk,gao,vicente,zhang}.

In this paper, we focus on the concurrence. We derive new lower and
upper bounds of concurrence for arbitrary quantum states. From the
bounds we can detect more entangled states. Detailed examples are given
to show that the new bounds of concurrence are better than that have
been obtained before.

For a pure N-partite quantum state $|\psi\ra\in {\mathcal
{H}}_{1}\otimes{\mathcal {H}}_{2}\otimes\cdots\otimes{\mathcal
{H}}_{N}$, $dim {\mathcal {H}}_{i}=d_i$, $i=1,...,N$, the concurrence of bipartite
decomposition between subsystems
$12\cdots M$ and $M+1\cdots N$ is defined by
\begin{eqnarray}\label{xx}
C_{2}(|\psi\ra\la\psi|)=\sqrt{2(1-{\rm Tr}\{\rho_{{1}{2}\cdots
{M}}^{2}\})}
\end{eqnarray}
where $\rho_{{1}{2}\cdots {M}}^{2}={\rm Tr}_{{M+1}\cdots
{N}}\{|\psi\ra\la\psi|\}$ is the reduced density matrix of
$\rho=|\psi\ra\la\psi|$ by tracing over subsystems $M+1\cdots{N}$.

On the other hand, the concurrence of $|\psi\ra$ is defined by \cite{multicon}
\begin{eqnarray}\label{xxx}
C_{N}(|\psi\ra\la\psi|)=2^{1-\frac{N}{2}}\sqrt{(2^{N}-2)-\sum_{\alpha}{\rm
Tr}\{\rho_{\alpha}^{2}\}},
\end{eqnarray}
where $\alpha$ labels all different reduced density matrices.

For a mixed multipartite quantum state, $\rho=\sum_{i}p_{i}|\psi_{i}\ra\la\psi_{i}|
\in {\mathcal {H}}_{1}\otimes{\mathcal {H}}_{2}\otimes\cdots\otimes{\mathcal {H}}_{N}$,
the corresponding concurrence of (\ref{xx}) and (\ref{xxx}) are then given by the convex roof:
\begin{eqnarray}\label{def1}
C_{2}(\rho)=\min_{\{p_{i},|\psi_{i}\}\ra}\sum_{i}p_{i}C_{2}(|\psi_{i}\ra\la\psi_{i}|),
\end{eqnarray}
\begin{eqnarray}\label{def}
C_{N}(\rho)=\min_{\{p_{i},|\psi_{i}\}\ra}\sum_{i}p_{i}C_{N}(|\psi_{i}\ra\la\psi_{i}|).
\end{eqnarray}
We now investigate the relation between the two kinds of
concurrences.

{\bf{Lemma 1:}} For a bipartite density matrix $\rho\in {\mathcal
{H}}_{A}\otimes{\mathcal {H}}_{B}$, one has
\begin{eqnarray}\label{yy}
1-{\rm Tr}\{\rho^{2}\}\leq 1-{\rm Tr}\{\rho_{A}^{2}\}+ 1-{\rm
Tr}\{\rho_{B}^{2}\},
\end{eqnarray}
where $\rho_{A/B}={\rm Tr}_{B/A}\{\rho\}$ be the reduced density
matrices.

{\bf{Proof:}} Let $\rho=\sum\limits_{ij}\lambda_{ij}|ij\ra\la ij|$
be the spectral decomposition, where $\lambda_{ij}\geq 0,
\sum_{ij}\lambda_{ij}=1$. Then
$\rho_{1}=\sum_{ij}\lambda_{ij}|i\ra\la i|,
\rho_{2}=\sum_{ij}\lambda_{ij}|j\ra\la j|$. Therefore
\begin{eqnarray*}
&&1-{\rm Tr}\{\rho_{A}^{2}\}+1-{\rm Tr}\{\rho_{B}^{2}\}-1+{\rm
Tr}\{\rho^{2}\}
=1-{\rm Tr}\{\rho_{A}^{2}\}-{\rm Tr}\{\rho_{B}^{2}\}+{\rm Tr}\{\rho^{2}\}\\
&=&(\sum_{ij}\lambda_{ij})^{2}-\sum_{i,j,j^{'}}\lambda_{ij}\lambda_{ij^{'}}
-\sum_{i,i^{'},j}\lambda_{ij}\lambda_{i^{'}j}+\sum_{ij}\lambda_{ij}^{2}\\
&=&(\sum_{i=i^{'},j=j^{'}}\lambda_{ij}^{2}+\sum_{i=i^{'},j\neq
j^{'}}\lambda_{ij}\lambda_{ij^{'}}+\sum_{i\neq i^{'},j=
j^{'}}\lambda_{ij}\lambda_{i^{'}j}+\sum_{i\neq i^{'},j\neq
j^{'}}\lambda_{ij}\lambda_{i^{'}j^{'}})-(\sum_{i,j=j^{'}}\lambda_{ij}^{2}+
\sum_{i,j\neq
j^{'}}\lambda_{ij}\lambda_{ij^{'}})\\
&&-(\sum_{i=i^{'},j}\lambda_{ij}^{2}+ \sum_{i\neq
i^{'},j}\lambda_{ij}\lambda_{i^{'}j})+\sum_{i,j}\lambda_{ij}^{2}\\
&=&\sum_{i\neq i^{'},j\neq
j^{'}}\lambda_{ij}\lambda_{i^{'}j^{'}}\geq 0.
\end{eqnarray*}
$\hfill\Box$

The same result in this lemma has also been derived in
\cite{zhang,cai} to prove the subadditivity of the linear entropy.
Here we just give a simpler proof. In the following we compare the
bi- and multi-partite concurrence in $(\ref{def1})(\ref{def})$ by
using the lemma.

{\bf{Theorem 1:}} For a multipartite quantum state $\rho\in
{\mathcal {H}}_{1}\otimes{\mathcal
{H}}_{2}\otimes\cdots\otimes{\mathcal {H}}_{N}$ with $N\geq 3$, the
following inequality holds,
\begin{eqnarray}
C_{N}(\rho)\geq\max 2^{\frac{3-N}{2}}C_{2}(\rho),
\end{eqnarray}
where the maximum is taken over all kinds of bipartite concurrence.

{\bf{Proof:}} Without lose of generality, we suppose that the
maximal bipartite concurrence is attained between subsystems
$12\cdots M$ and $(M+1)\cdots N$.

For a pure multipartite state $|\psi\ra\in {\mathcal
{H}}_{1}\otimes{\mathcal {H}}_{2}\otimes\cdots\otimes{\mathcal
{H}}_{N}$, ${\rm Tr}\{\rho_{12\cdots M}^{2}\}={\rm
Tr}\{\rho_{(M+1)\cdots N}^{2}\}$. From (\ref{yy}) we have
\begin{eqnarray*}
C_{N}^{2}(|\psi\ra\la\psi|)&=&2^{2-N}((2^{N}-2)-\sum_{\alpha}{\rm
Tr}\{\rho_{\alpha}^{2}\})\geq
2^{3-N}(N-\sum_{k=1}^{N}{\rm Tr}\{\rho_{k}^{2}\})\\
&\geq& 2^{3-N}(1-{\rm Tr}\{\rho_{12\cdots M}^{2}\}+1-{\rm
Tr}\{\rho_{(M+1)\cdots
N}^{2}\})\\
&=&2^{3-N}*2(1-{\rm Tr}\{\rho_{12\cdots
M}^{2}\})=2^{3-N}C_{2}^{2}(|\psi\ra\la\psi|),
\end{eqnarray*}
i.e. $C_{N}(|\psi\ra\la\psi|)\geq
2^{\frac{3-N}{2}}C_{2}(|\psi\ra\la\psi|)$.

Let $\rho=\sum\limits_{i}p_{i}|\psi_{i}\ra\la\psi_{i}|$ attain the
minimal decomposition of the multipartite concurrence. One has
\begin{eqnarray*}
C_{N}(\rho)=\sum_{i}p_{i}C_{N}(|\psi_{i}\ra\la\psi_{i}|)\geq
2^{\frac{3-N}{2}}\sum_{i}p_{i}C_{2}(|\psi_{i}\ra\la\psi_{i}|)\\
\geq2^{\frac{3-N}{2}}\min_{\{p_{i},|\psi_{i}\}}
\sum_{i}p_{i}C_{2}(|\psi_{i}\ra\la\psi_{i}|)=2^{\frac{3-N}{2}}C_{2}(\rho).
\end{eqnarray*}
$\hfill\Box$

{\bf{Corollary}} For a tripartite quantum state $\rho\in{\mathcal
{H}}_{1}\otimes{\mathcal {H}}_{2}\otimes{\mathcal {H}}_{3}$, the
following inequality hold:
\begin{eqnarray}
C_{3}(\rho)\geq\max C_{2}(\rho)
\end{eqnarray}
where the maximum is taken over all kinds of bipartite concurrence.

In \cite{chenk} a lower bound for a bipartite state $\rho\in
{\mathcal {H}}_{A}\otimes{\mathcal {H}}_{B}$, $d_A\leq d_B$, has
been obtained,
\begin{eqnarray}\label{lowerbound1}
C_{2}(\rho)\geq\sqrt{\frac{2}{d_A(d_A-1)}}[\max(||{\mathcal
{T}}_{A}(\rho)||,||R(\rho)||)-1].
\end{eqnarray}
where ${\mathcal {T}}_{A}$, $R$ and $||\cdot||$ stand for the
partial transpose, realignment, and the trace norm (i.e., the sum of
the singular values), respectively.

In \cite{vicente,zhang1}, from the separability criteria related to
local uncertainty relation, covariance matrix and
correlation matrix, the following lower bounds for bipartite concurrence are obtained:
\begin{eqnarray}\label{lowerbound2}
C_{2}(\rho)\geq\frac{2||C(\rho)||-(1-{\rm
Tr}\{\rho_{A}^{2}\})-(1-{\rm
Tr}\{\rho_{B}^{2}\})}{\sqrt{2d_A(d_A-1)}}
\end{eqnarray}
and
\begin{eqnarray}\label{lowerbound3}
C_{2}(\rho)\geq\sqrt{\frac{8}{d_A^{3}d_B^{2}(d_A-1)}}(||T(\rho)||-\frac{\sqrt{d_Ad_B(d_A-1)(d_B-1)}}{2}),
\end{eqnarray}
where the entries of the matrix $C$,
$C_{ij}=\la\lambda^{A}_{i}\otimes\lambda^{B}_{j}\ra-\la\lambda^{A}_{i}\otimes
I_{d_B}\ra\la I_{d_A}\otimes\lambda^{B}_{j}\ra$,
$T_{ij}=\frac{d_Ad_B}{2}\la\lambda^{A}_{i}\otimes\lambda^{B}_{j}\ra$,
$\lambda^{A/B}_{k}$ stands for the normalized generator of
$SU(d_A/d_B)$, i.e. ${\rm
Tr}\{\lambda^{A/B}_{k}\lambda^{A/B}_{l}\}=\delta_{kl}$ and $\la
X\ra={\rm Tr}\{\rho X\}$. It is shown that the lower bounds
$(\ref{lowerbound2})$ and $(\ref{lowerbound3})$ are independent of
$(\ref{lowerbound1})$.

Now we consider a multipartite quantum state $\rho\in{\mathcal
{H}}_{1}\otimes{\mathcal {H}}_{2}\otimes\cdots\otimes{\mathcal
{H}}_{N}$ as a bipartite state belonging to ${\mathcal
{H}}^{A}\otimes{\mathcal {H}}^{B}$ with the dimensions of the
subsystems A and B being $d_A=d_{s_{1}}d_{s_{2}}\cdots d_{s_{m}}$
and $d_B=d_{s_{m+1}}d_{s_{m+2}}\cdots d_{s_{N}}$ respectively. By
using the corollary, $(\ref{lowerbound1})$, $(\ref{lowerbound2})$
and $(\ref{lowerbound3})$ we have the following lower bound:

{\bf{Theorem 2:}} For any N-partite quantum state $\rho$, we have:
\begin{eqnarray}\label{newlowerbound}
C_{N}(\rho)\geq2^{\frac{3-N}{2}}\max\{B1,B2,B3\},
\end{eqnarray}
where
\begin{eqnarray*}
B1&=&\max_{\{i\}}\sqrt{\frac{2}{M_{i}(M_{i}-1)}}\left[\max(||{\mathcal
{T}}_{A}(\rho^{i})||,||R(\rho^{i})||)-1\right],\\
B2&=&\max_{\{i\}}\frac{2||C(\rho^{i})||- (1-{\rm
Tr}\{(\rho^{i}_{A})^{2}\})-(1-{\rm Tr}\{(\rho^{i}_{B})^{2}\})}
{\sqrt{2M_{i}(M_{i}-1)}},\\
B3&=&\max_{\{i\}}\sqrt{\frac{8}{M_{i}^{3}N_{i}^{2}(M_{i}-1)}}
(||T(\rho^{i})||-\frac{\sqrt{M_{i}N_{i}(M_{i}-1)(N_{i}-1)}}{2}),
\end{eqnarray*}
$\rho^i$s are all possible bipartite decompositions of $\rho$, and
$M_{i}=\min{\{d_{s_{1}}d_{s_{2}}\cdots d_{s_{m}},
d_{s_{m+1}}d_{s_{m+2}}\cdots d_{s_{N}}\}}$,
$N_{i}=\max{\{d_{s_{1}}d_{s_{2}}\cdots d_{s_{m}},
d_{s_{m+1}}d_{s_{m+2}}\cdots d_{s_{N}}\}}$.

In \cite{zhang,mintert,aolita}, it is shown that the upper and
lower bound of multipartite concurrence satisfy
\begin{eqnarray}\label{upperlowerbound}
\sqrt{(4-2^{3-N}){\rm Tr}\{\rho^{2}\}-2^{2-N}\sum_{\alpha}{\rm
Tr}\{\rho_{\alpha}^{2}\}}\leq
C_{N}(\rho)\leq\sqrt{2^{2-N}[(2^{N}-2)-\sum_{\alpha}{\rm
Tr}\{\rho_{\alpha}^{2}\}]}.
\end{eqnarray}

In fact we can obtain a more effective upper bound for
multi-partite concurrence. Let
$\rho=\sum\limits_{i}\lambda_{i}|\psi_{i}\ra\la \psi_{i}|\in
{\mathcal {H}}_{1}\otimes{\mathcal{H}}_{2}\otimes\cdots\otimes{\mathcal {H}}_{N}$, where
$|\psi_{i}\ra$s are the  orthogonal pure states and
$\sum\limits_{i}\lambda_{i}=1$. We have
\begin{eqnarray}\label{newupperbound}
C_{N}(\rho)=\min_{\{p_{i},|\varphi_{i}\}\ra}\sum_{i}p_{i}C_{N}(|\varphi_{i}\ra\la\varphi_{i}|)
\leq\sum_{i}\lambda_{i}C_{N}(|\psi_{i}\ra\la\psi_{i}|).
\end{eqnarray}
The right side of $(\ref{newupperbound})$ gives a new upper bound of $C_{N}(\rho)$. Since
\begin{eqnarray*}
\sum_{i}\lambda_{i}C_{N}(|\psi_{i}\ra\la\psi_{i}|)
&=&2^{1-\frac{N}{2}}\sum_{i}\lambda_{i}\sqrt{(2^{N}-2)-\sum_{\alpha}{\rm Tr}\{(\rho^{i}_{\alpha})^{2}\}}\\
&\leq&
2^{1-\frac{N}{2}}\sqrt{(2^{N}-2)-\sum_{\alpha}{\rm Tr}\{\sum_{i}\lambda_{i}(\rho^{i}_{\alpha})^{2}\}}\\
&\leq& 2^{1-\frac{N}{2}}\sqrt{(2^{N}-2)-\sum_{\alpha}{\rm
Tr}\{(\rho_{\alpha})^{2}\}},
\end{eqnarray*}
the upper bound obtained in $(\ref{newupperbound})$ is
better than that in $(\ref{upperlowerbound})$.

The lower and upper bounds can be used to estimate the value of the
concurrence. Meanwhile, the lower bound of concurrence can be used
to detect entanglement of quantum states. We now show that our upper
and lower bounds can be better than that in
$(\ref{upperlowerbound})$ by several detailed examples.

\begin{figure}[tbp]
\begin{center}
\resizebox{10cm}{!}{\includegraphics{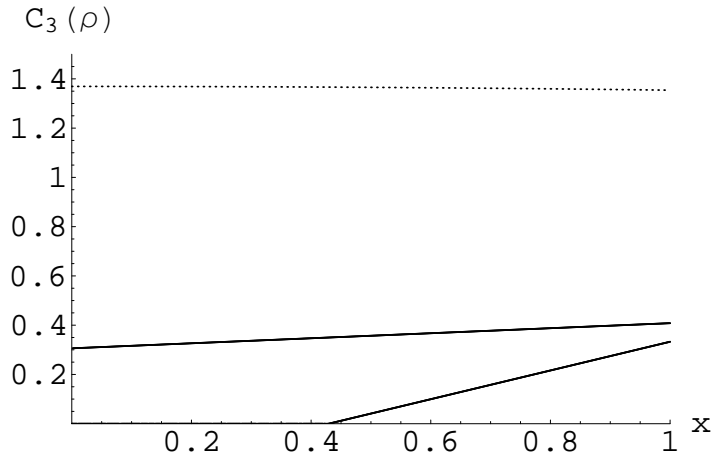}}
\end{center}
\caption{Our lower and upper bounds of $C_{3}(\rho)$ from
$(\ref{newlowerbound})(\ref{newupperbound})$(solid line) and the
upper bound obtained in $(\ref{upperlowerbound})$(dot line) while
the lower bound in $(\ref{upperlowerbound})$ is always zero.
\label{fig1}}
\end{figure}

{\bf{Example 1:}} Consider the $2\times 2\times 2$ D\"u
r-Cirac-Tarrach states defined by \cite{dur}:
\begin{eqnarray}
\rho=\sum_{\sigma=\pm}\lambda_{0}^{\sigma}|\Psi_{0}^{\sigma}\ra\la
\Psi_{0}^{\sigma}|+\sum_{j=1}^{3}\lambda_{j}(|\Psi_{j}^{+}\ra\la\Psi_{j}^{+}|+|\Psi_{j}^{-}\ra\la\Psi_{j}^{-}|),
\end{eqnarray}
where the orthonormal Greenberger-Horne-Zeilinger (GHZ)-basis
$|\Psi_{j}^{\pm}\ra\equiv\frac{1}{\sqrt{2}}(|j\ra_{12}|0\ra_{3}\pm|(3-j)\ra_{12}|1\ra_{3}),
|j\ra_{12}\equiv|j_{1}\ra_{1}|j_{2}\ra_{2}$ with $j=j_{1}j_{2}$ in
binary notation. From theorem 2 we have that the
lower bound of $\rho$ is $\frac{1}{3}$. If we mix the state with white noise,
\begin{eqnarray}
\rho(x)=\frac{(1-x)}{8}I_{8}+x\rho,
\end{eqnarray}
by direct computation we have, as shown in FIG. $\ref{fig1}$, the
lower bound obtained in $(\ref{upperlowerbound})$ is always zero,
while the lower bound in $(\ref{newlowerbound})$ is larger than zero
for $0.425\leq x\leq 1$, which shows that $\rho(x)$ is detected to
be entangled at this situation. And the upper bound (dot line) in
$(\ref{upperlowerbound})$ is much larger than the upper bound we
have obtained in $(\ref{newupperbound})$ (solid line).

\begin{figure}[h]
\begin{center}
\resizebox{10cm}{!}{\includegraphics{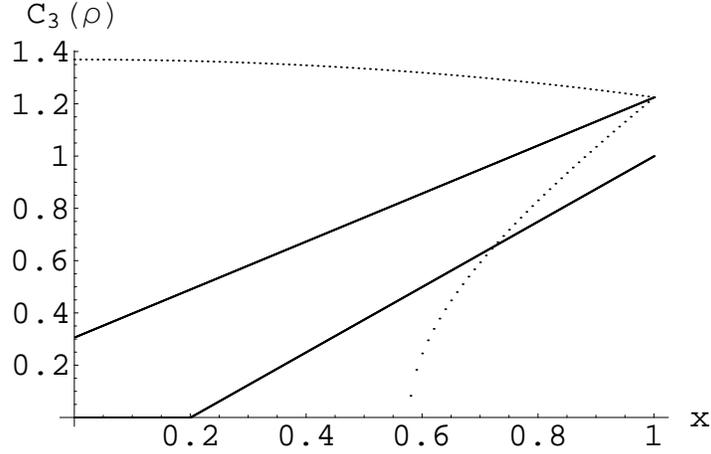}}
\end{center}
\caption{Our lower and upper bounds of $C_{3}(\rho)$ from
$(\ref{newlowerbound})$$(\ref{newupperbound})$ (solid line) and the
bounds obtained in $(\ref{upperlowerbound})$(dot line).
\label{fig2}}
\end{figure}

{\bf{Example 2:}} We consider the depolarized state \cite{dur}:
\begin{eqnarray}
\rho=\frac{(1-x)}{8}I_{8}+x|\psi^{+}\ra\la\psi^{+}|,
\end{eqnarray}
where $0\leq x\leq 1$ representing the degree of depolarization,
$|\psi^{+}\ra=\frac{1}{\sqrt{2}}(|000\ra+|111\ra)$. From FIG.
$\ref{fig2}$ one can obviously seen that our upper bound is tighter.
For $0\leq x\leq 0.7237$ our lower bound is higher than that in
$(\ref{upperlowerbound})$, i.e. our lower bound is closer to the
true concurrence. Moreover for $0.2\leq x\leq 0.57735$, our lower
bound can detect the entanglement of $\rho$, while the lower bound
in $(\ref{upperlowerbound})$ not.

\medskip
We have studied the concurrence for arbitrary multipartite quantum states. We
derived new better lower and upper bounds. The lower bound can also be used to detect more
multipartite entangled quantum states.

\bigskip
\noindent{\bf Acknowledgments}\, This work is supported by NSFC
under grant 10675086, NKBRSFC under grant 2004CB318000.

\end{document}